\documentclass{llncs}
\usepackage{epsfig,endnotes}
\usepackage{color}
\usepackage{url}
\usepackage{listings}
\usepackage{fancyvrb}
\usepackage{subfigure}
\usepackage[square,sort&compress,comma,numbers]{natbib}
\usepackage{algorithm}
\usepackage[noend]{algpseudocode}
\algnewcommand{\LineComment}[1]{\State \(\triangleright\) #1}
\usepackage{hyphenat}
\usepackage{pgfplots}
\usepackage{hyperref}
\usepackage{booktabs}

\newcommand{\code}[1]{\texttt{\small #1}}

\pgfplotsset{
  width=7cm
}

\begin{document}

\title{Leveraging Flawed Tutorials for Seeding Large-Scale Web Vulnerability Discovery}

\author{Tommi Unruh\inst{1} \and Bhargava Shastry\inst{1} \and
Malte Skoruppa\inst{2} \and \\
Federico Maggi\inst{3} \and Konrad Rieck\inst{4} \and Jean-Pierre Seifert\inst{1} \and \\
Fabian Yamaguchi\inst{4}
}
\institute{
Security in Telecommunications, TU Berlin, Germany
\and
CISPA, Saarland University
\and
Trend Micro Inc.
\and
Institute of System Security, TU Braunschweig, Germany
}

\maketitle
\begin{abstract}
The Web is replete with tutorial\hyp{}style content on how to accomplish programming tasks.
Unfortunately, even top\hyp{}ranked tutorials suffer from severe security vulnerabilities, such as cross\hyp{}site scripting (XSS), and SQL injection (SQLi).
Assuming that these tutorials influence real\hyp{}world software development, we hypothesize that code snippets from popular tutorials can be used to bootstrap vulnerability discovery at scale.
To validate our hypothesis, we propose a semi\hyp{}automated approach to find recurring vulnerabilities starting from a handful of top\hyp{}ranked tutorials that contain vulnerable code snippets.
We evaluate our approach by performing an analysis of tens of thousands of open\hyp{}source web applications to check if vulnerabilities originating in the selected tutorials recur.
%we perform an analysis of tens of thousands of web applications to check if vulnerabilities originating in top\hyp{}ranked tutorials recur.
Our analysis framework has been running on a standard PC, analyzed 64,415 PHP codebases hosted on \emph{GitHub} thus far, and found a total of 117 vulnerabilities that have a strong syntactic similarity to vulnerable code snippets present in popular tutorials.
In addition to shedding light on the anecdotal belief that programmers reuse web tutorial code in an ad hoc manner, our study finds disconcerting evidence of insufficiently reviewed tutorials compromising the security of open\hyp{}source projects.
Moreover, our findings testify to the feasibility of large\hyp{}scale vulnerability discovery using poorly written tutorials as a starting point.
\end{abstract}

% \keywords{ACM proceedings; \LaTeX; text tagging}

\section{Introduction}

Programming aids such as web tutorials and Q\&A websites are popular
with novice and expert programmers alike. To what extent, and how
these aids influence the quality of real\hyp{}world software remains
an open question.  On the one hand, popular Q\&A websites, such as
{\tt stackoverflow.com}, have an in\hyp{}built reputation system
where correct advice gets up-voted through a consensus.  Hence, a
common expectation is that community\hyp{}driven websites weed out bad
coding suggestions. On the other hand, the Web is replete with
tutorial\hyp{}style webpages that simply present curated snippets of
code that accomplish a task.  Most tutorials omit a discussion about
API quirks or the security\hyp{}impact a code snippet might have. It
is not surprising, therefore, that the presented code snippets suffer
from basic security vulnerabilities.

The connection between tutorials and vulnerabilities in
real\hyp{}world code is largely unknown.  Although previous
studies~\citep{Jang2012,Pham2010, GauLavMer13, Yamaguchi2012} have
shown that copy\hyp{}pasted code can lead to recurring
vulnerabilities, these studies have only considered instances of
copy\hyp{}pasted code \emph{within} a codebase.  We seek to generalize
this result by asserting that, like code snippets originating in the
same codebase, popular programming resources on the Web constitute a major source
of documentation that is regularly consulted by developers
and often introduces vulnerabilities into software.

Based on our assertion, we hypothesize that vulnerability discovery
can be \emph{seeded} by code snippets such as those found in top\hyp{}ranked
tutorials.  Viewed from an adversarial standpoint, we present a novel
approach for bootstrapping vulnerability discovery at scale.  Our main
intuition is that recurring vulnerabilities can be found by
recognizing, and subsequently looking for patterns in code that
correspond to the original vulnerability. We refer to instances of these patterns
as \emph{code analogues} throughout the rest of the paper. Our
expectation is that if such a pattern recurs, so will the
corresponding vulnerability. To identify code analogues, we
automatically generate \emph{graph traversals}, which can be used to
mine code for these analogues using graph databases. Each graph
traversal is derived from normalized fragments of a code snippet's
abstract syntax tree (AST), augmented with data-flow
information. These graph traversals thus express syntactic properties
of the original tutorial code.

\begin{figure*}[t]
  \centering
  \includegraphics[scale=.57]{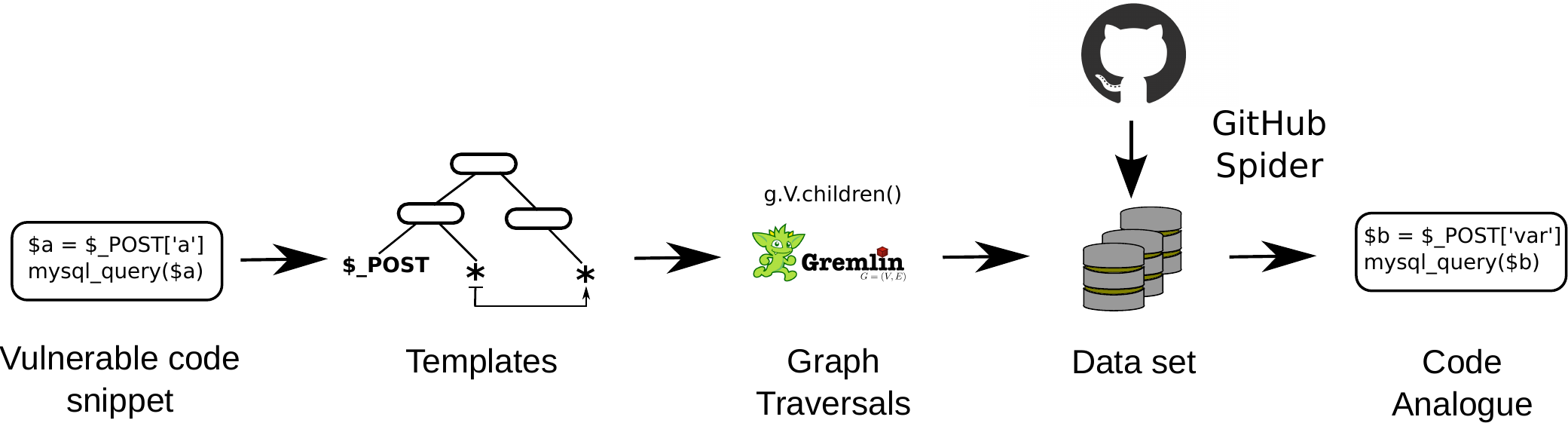}
  \caption{Workflow for finding recurring vulnerabilities.}
  \label{fig:workflow}
\end{figure*}

Our workflow for finding recurring vulnerabilities consists of two
steps. First, we automatically translate vulnerable code snippets into
graph traversals, which concretize our expectations of code that
constitutes an analogue. Second, we serialize a codebase under
analysis into a graph database and employ the automatically generated
traversals to search the database for analogue occurrences. As a
result, we obtain a set of locations in application source code that bear a strong syntactic resemblance to vulnerable code snippets.

% We refer to
% syntactically similar code as \emph{code analogues}, or simply
% analogues, throughout the rest of this paper.

An empirical evaluation of our approach, on a data set of 64,415
web applications, shows that an adversary with access to a standard PC
and a DSL broadband connection can leverage our techniques to
efficiently discover recurring vulnerabilities in web application code
(Section \ref{sec:eval}). Although AST is a fundamental construct for any programming language, we prototype our analysis framework for analyzing applications written in PHP, the most widely deployed server side scripting language to date~\citep{vuln:weblanguages_distribution}. 
Our analysis framework
accepts vulnerable (tutorial) code snippets as input, and returns its analogues
in a target PHP codebase.  Since our analysis can not guarantee that
the returned analogues are also vulnerable, we manually review them.
Manual review has also been useful in teasing out the connection
insecure tutorials and web application vulnerabilities. Thanks to our framework, we
have uncovered over 100 vulnerabilities in web application code that
bear a strong resemblance to vulnerable code patterns found in popular
tutorials.  More alarmingly, we have confirmed that 8 instances of a
SQLi vulnerability present in different web applications are an
outcome of code copied from a single vulnerable tutorial.
%Although our results do not establish a causal relationship between insecure tutorials and web application vulnerabilities, we find the
Our results indicate that there is a substantial, if not causal, link between insecure tutorials and web application vulnerabilities.

% that 8 instances of a SQLi vulnerability present in multiple distinct
% uncovered 8 instances of a SQLi vulnerability spread across distinct web applications that is an outcome of code copied from a single vulnerable tutorial.
%  SQLi vulnerabilities in web application code
% Thus far, our analysis has uncovered over 100 vulnerabilities in application code that have a syntactic resemblance to similar vulnerabilities in popular tutorials.
% During manual validation, we have found that 8 of the uncovered vulnerabilities are
% the sheer number of recurring vulnerabilities that it has helped discover, indicates that there is a substantial link between insecure tutorials and web application vulnerabilities.

In summary, we make the following contributions:
\begin{itemize}
\item We present a novel approach for bootstrapping large-scale
  vulnerability discovery, namely, leveraging flawed tutorial code to seed vulnerability search in application code. We evaluate this approach using a proof-of-concept framework that finds recurring vulnerabilities in PHP application code hosted on GitHub.

\item We propose a computationally efficient method to search for
  recurring vulnerabilities. We translate vulnerable snippets of code
  into graph traversals to identify code analogues in a program
  abstraction.

\item We show that large-scale vulnerability scanning of publicly
  available open-source repositories is feasible, even with limited
  resources such as a standard PC and a broadband DSL
  connection.

\item Finally, our results give credence to the widely known anecdote that
  programmers copy and paste code from vulnerable tutorials. Our case study, involving 64,415 PHP projects hosted on GitHub, indicates that such ad hoc code re\hyp{}use may endanger the security of software throughout the open\hyp{}source landscape.
\end{itemize}

Our tools GithubSpider, and CADetector are available at \url{https://github.com/tommiu/GithubSpider} and \url{https://github.com/tommiu/ccdetection} respectively.

% The remainder of this paper is organized as follows. We describe our approach for identifying tutorial-induced
% vulnerabilities in Section~\ref{sec:methodology}. We evaluate our
% approach in Section~\ref{sec:eval}, and discuss related work and
% limitations in Section~\ref{sec:relatedwork} and
% Section~\ref{sec:limitations}
% respectively. Section~\ref{sec:conclusion} concludes the paper.

\section{Methodology}
\label{sec:methodology}

In this study, we take
on an attacker's perspective and design our method such that it allows
for semi-automated discovery of recurring vulnerabilities without
requiring special access to a hosting platform or considerable
computational resources.
Figure~\ref{fig:workflow} illustrates our workflow, that we briefly describe in the following paragraphs.

\begin{enumerate}

% \item \textbf{Finding vulnerabilities in tutorials.} We proceed to
%   search the Web for popular tutorials and analyze them for
%   vulnerabilities by following OWASP guides on the prevention of cross
%   site scripting and SQL-injection vulnerabilities.

\item \textbf{Derivation of templates from tutorials.} We extract templates from vulnerable code snippets contained in
  tutorials. These templates represent syntactical properties of the
  vulnerable code as well as information about data flow
  (Section~\ref{sec:template}).

\item \textbf{Generating traversals from templates.} We leverage templates to automatically generate \emph{traversals} for
  a graph-based code mining system. Graph traversals enable us to scan
  large amounts of code for analogues of these vulnerable snippets in a computationally efficient manner (Section~\ref{sec:querygen}).

\item \textbf{Spidering code repositories.} We automatically collect a large data set of
  open-source code bases from a code hosting site, choosing GitHub as a representative case study (Section~\ref{sec:spidering}).

\item \textbf{Mining for vulnerabilities.} Leveraging our analysis framework, we automatically mine the code
  of our data set for instances of vulnerable tutorial snippets.
  We manually cross\hyp{}check if matches returned by our analysis platform constitute vulnerabilities (Section~\ref{sec:mining}).

\end{enumerate}

\noindent{}In the remainder of this section, we describe each of these steps in greater detail providing background information where necessary.

\subsection{Derivation of Templates from Tutorials}
\label{sec:template}

Exact copies of vulnerable code snippets can be found using string
matching utilities such as the standard UNIX tool
{\tt grep}. However, as programmers copy and paste code from
tutorials, they are likely to adapt it slightly, for example, by
changing the names of variables. Therefore this na{\"i}ve approach fails
in all but the most simple cases. To account for slight modifications,
we require a method that is robust enough to identify sequences of statements \textit{similar} to those found in the tutorial in terms of the operations they carry out on their input variables. We refer to these re-occurrences of tutorial
code as \emph{code analogues}, or simply analogues.

An elegant approach to address this problem is to extract intermediate
graph representations from code that represent syntax and data flow,
and formulate syntactical and data flow properties of the code snippet
in terms of traversals in these graphs~\citep[see][]{Rep98,
Yamaguchi14, he2015vetting}. These traversals are formulated such that
they succeed when the code matches, and fail when it does not. Although
these graph traversals can be formulated manually, in this work, we
devise a two-step procedure to automatically generate them from
vulnerable snippets of code, making it possible to directly search for
these snippets without additional manual work.

\begin{figure}[t]
  \begin{minipage}[t]{\columnwidth}
\begin{Verbatim}
<?php
 include "db.php";

 @hili`$title=$_POST["title"];|
 @hili`$result=mysql_query("SELECT * FROM wp_posts where;|
    @hili`post_title like '%$title%' and post_status='publish'");|
 @hili`$found=mysql_num_rows($result);|

 if($found>0){
        while($row=mysql_fetch_array($result)){
        echo "<li>$row[post_title]</br>
                <a href=$row[guid]>$row[guid]</a></li>";
    }
 }else{
        echo "<li>No Tutorial Found<li>";
 }
 // ajax search
?>
\end{Verbatim}
   \end{minipage}
   \caption{Identified vulnerable tutorial, allowing for SQLi (line
     6), and XSS (line 11-12).}
   \label{fig:sampleTutorial}
 \end{figure}

The first step of our procedure is to generate a \emph{template} that
encodes syntax and data flow of the code snippet that we attempt to scan
for. To illustrate this process, we consider the vulnerable code
snippet shown in Figure~\ref{fig:sampleTutorial} taken from a popular
PHP tutorial. The code contains one SQLi, and one stored XSS vulnerability. The SQLi vulnerability occurs on
line $6$ as the attacker-controlled POST-variable \code{\$title} is
used in the construction of an SQL query without first undergoing
sanitization. The XSS vulnerability can be triggered on line $11$, and $12$ where databases rows are inserted into the document without escaping.

Two queries can be generated from the code snippet shown in Figure~\ref{fig:sampleTutorial}: one to identify
instances of the SQLi vulnerability, and another for the
XSS vulnerability. In the following walk\hyp{}through, we focus on the
SQLi vulnerability, as highlighted in Figure~\ref{fig:sampleTutorial}.

\begin{figure*}[t]
  \centering
  \subfigure[Abstract Syntax Tree.]{\label{fig:ast}
    \includegraphics[scale=.9]{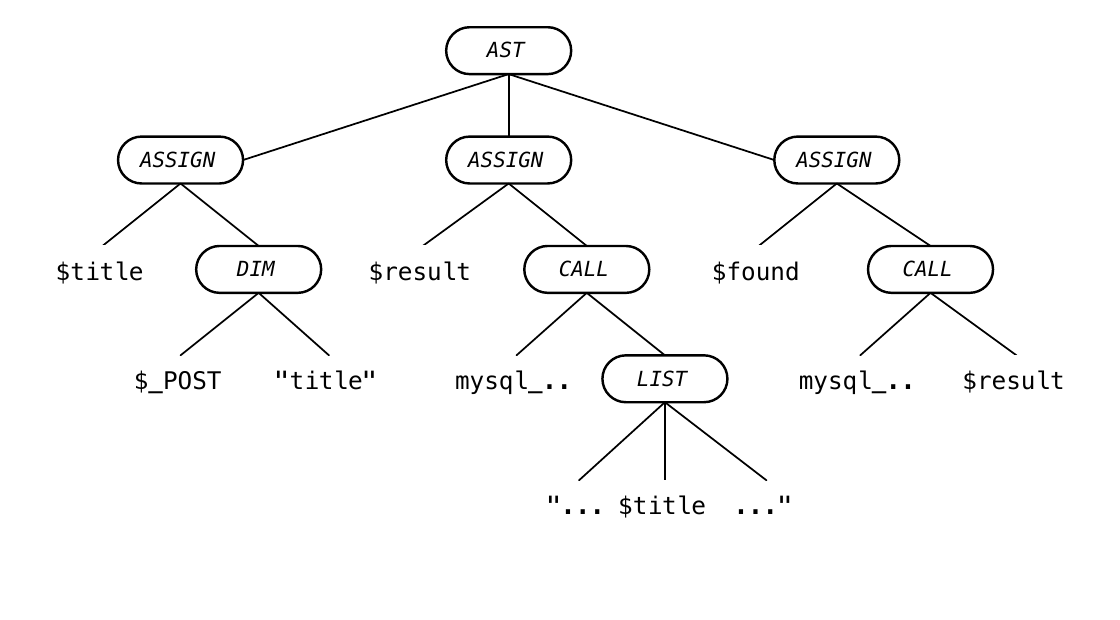}} \quad
  \subfigure[Derived Template.]{\label{fig:template}
     \includegraphics[scale=.9]{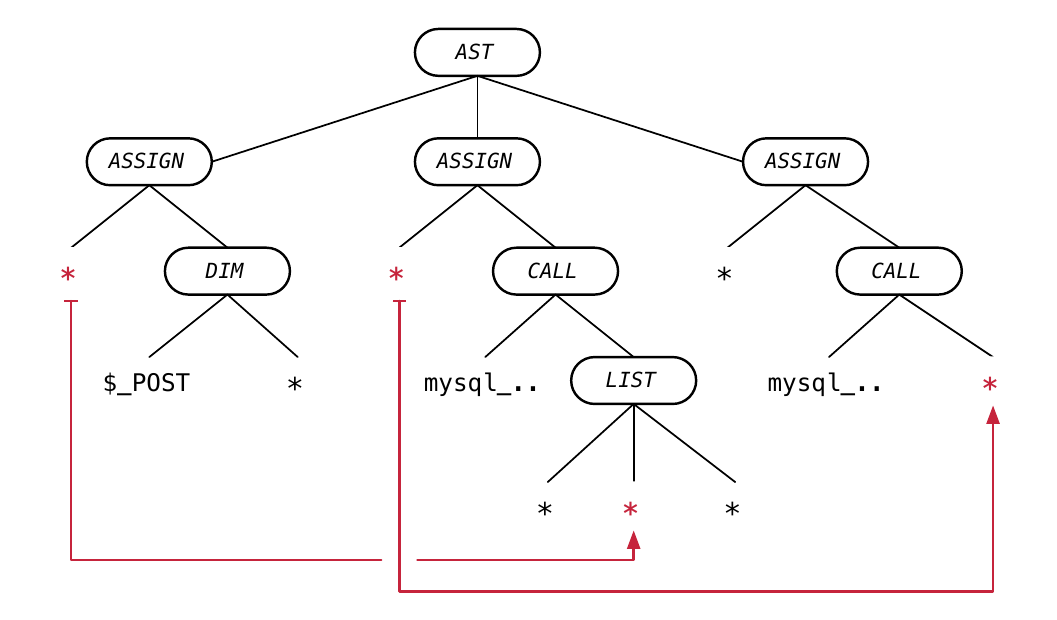}}
  \caption{Abstract syntax tree and derived template for
  the example in Figure~\ref{fig:sampleTutorial}.}
  \label{fig:ast_tmp}
\end{figure*}

We proceed to generate an AST of the vulnerable code snippet, a standard tree-representation of program syntax. ASTs provide a hierarchical decomposition of code into
its language elements.
As an example, Figure~\ref{fig:ast} shows the abstract syntax tree for
the SQLi vulnerability. In this tree, leaf nodes correspond
to identifiers (e.g., \code{\$title}), API symbols (e.g.,
\code{mysql\_...}), or literals (e.g., \code{SELECT}), and inner
nodes represent operations such as assignments, function calls, or
array indexing operations.

To derive a template from an AST, we replace all
variables and literals by wildcard symbols and introduce edges between
nodes representing the same variable. The template thus abstracts from
concrete variable names and strings, while preserving data flow between
variables. Figure~\ref{fig:template} shows the corresponding template
for our running example. As indicated by the data flow edges, the
template enforces that there is a match if and only if the variable
occurring on the left-hand side of the assignment in the first
statement, and the variable appearing as an argument in a function
call in the next statement are the same. An assignment to one
variable, followed by the use of a different variable in a function
call does not trigger a match, because there is no data flow between them.

With templates for code snippets at hand, we are now ready to generate
graph traversals that allow code to be mined for instances (analogues) of these
snippets.

% \begin{itemize}

% % \item Grep has no understanding of the code's syntax. We cannot deal
% %   with changes of variable names. However, when copying code from a
% %   tutorial, it is likely that the programmer changes the names of.

% % \item Looking for structural clones: consecutive lines of code that
% %   treat variables in the same way as the code snippet.

%   \item What the algorithm does, is to look at each child of the AST
%     root and create traversals that guarantee, that these child nodes
%     are in correct order and have the correct node types.

%   \item Syntax-awareness: For a given string, syntax-awareness allows
%     us to see whether a string is a literal or, e.g., represents the
%     name of a function.

%   \item Allows for normalization: literal strings and integers: ignore
%     values and convert them into general string or integer node
%     representations.

%   \item Build Gremlin queries that detect exact copies, ignoring
%     actual values of strings and integers, and modified copies of the
%     source AST.

%   \item Data flow: names of variables are taken into account.

%   % \item Main question: do we preserve the names of API symbols, e.g.,
%   %   mysql\_query? I think, we don't.: We don't.

% \end{itemize}

% Our approach extracts an abstract syntax tree that corresponds to the
% vulnerable code snippet, and subsequently translated this tree into a
% query. This query encodes the vulnerable code in terms of its
% syntactical properties, however, it also allows for normalization of
% literal strings and numbers.

\subsection{Generation of Traversals from Templates}
\label{sec:querygen}

Upon successful generation of a template for a vulnerable code
snippet, we transform the template into a corresponding graph
traversal. Although in principle, graph traversals can be formulated in
any programming language, the query language
Gremlin~\citep{gremlin:docs} is specifically designed for this
purpose. Moreover, traversals formulated in Gremlin can be executed on
any graph database system that supports the Blueprints
standard~\citep{tinkerpop:blueprints_model}, a compatibility layer for
graph databases similar to JDBC for relational databases.

For a given template, we generate a traversal that identifies
ASTs with (1) the same node types, (2) in the same order, and
(3) nested in the same way. Moreover, the traversal succeeds only if data
flow between statements is in correspondence with the data-flow edges
of the template.

In essence, the AST structure is encoded by formulating chains of filter
operations that succeed only if the desired node types can be
matched. To account for data flow, the names of variables are stored
as the AST is traversed, and filtering is performed to ensure
that the correct variable names occur throughout the
tree. Algorithm~\ref{php2gremlin:construct_query} describes this
process in detail. Each child node of the template is
converted into a traversal via the recursive function
\texttt{ConvertNode}, and \texttt{prepareNode} adds code
to traverse to the next child node.

As an example, Figure~\ref{fig:traversal} shows the traversal
generated by this procedure for the vulnerable code snippet in
Figure~\ref{fig:sampleTutorial}. For all nodes in the AST, the
traversal attempts to match the left subtree, the subtree in the middle,
and the right subtree starting at lines $3$, $20$, and $37$
respectively. Subtrees are matched by applying chains of filter
expressions. For instance, to match the left-most subtree consisting of an
assignment with a variable as a child node, the traversal first
attempts to match an assignment, and, on success, determines whether
the child node is a variable node using a subsequent filter (see lines
$4$ and $5$). Finally, lightweight data-flow tracking is implemented
by storing the names of variables as they are first encountered (e.g.,
on line $8$), and filtering based on these names (e.g., line $33$).

\begin{algorithm}[t]
\caption{Pseudocode for translating templates into graph traversals.}
\label{php2gremlin:construct_query}

\begin{algorithmic}[1]
\Function{traversal}{$template$}
\State t = $\emptyset$
% \State remember line number of first matched node
\For{each $child$ of $template$}
        \State t += convertNode($child$)
        \State t += prepareNextSubtree()
\EndFor
\Return t
% \State remember line number of last matched node
% \State print code clones
\EndFunction
\\
\Function{convertNode}{$child$}
\State t = $\emptyset$
\LineComment Ensure correct node-type
\State t+= filter\{ "node.type == child.type" \}
\If{node.type == "VAR"}
        \If{variable with $name$ has been used before}
                \State t += filter \{ "this\_node's name == $name$" \}
        \Else
                \State t += sideEffect \{ Remember the variable \}
        \EndIf
\EndIf

\For{each $child\_node$ of $child$}
        \State t += convertNode(child\_node)
\EndFor
\Return t
\EndFunction
\\
\Function{prepareNextSubtree}{}
\State
\Return \{ traverse next child \}
\EndFunction
\end{algorithmic}
\end{algorithm}

\begin{figure}[tbh]
  \begin{minipage}[t]{\columnwidth}
\begin{Verbatim}
g.V.sideEffect{ childnumber = 0 }

// Left subtree
.filter{ isType(it, "ASSIGN") }
.filter{ isType(it.ithChildren(0), "VAR") }

// Remember variable "title"
.sideEffect{ _title = it.ithChildren(0).varToName().next() }
.filter{ isType(it.ithChildren(1), "DIM") }
.filter{ isType(it.ithChildren(1).ithChildren(0), "VAR") }
.sideEffect{ childnumber = it.childnum }
.sideEffect{ childnumber = childnumber + 1 }

.parents().children().filter{ it.childnum == childnumber }

// Subtree in the middle
.filter{ isType(it, "ASSIGN") }
.filter{ isType(it.ithChildren(0), "VAR") }

// Remember variable "result"
.sideEffect{ _result = it.ithChildren(0).varToName().next() }
.filter{ isType(it.ithChildren(1), "CALL") }
.filter{ isType(it.ithChildren(1)
  .ithChildren(0), "NAME") }
.filter{ isType(it.ithChildren(1)
  .ithChildren(1), "ARG_LIST") }
...

.sideEffect{ childnumber = it.childnum }
.sideEffect{ childnumber = childnumber + 1 }

// Right subtree
.parents().children()
.filter{ it.childnum == childnumber }
.filter{ isType(it, "ASSIGN") }
.filter{ isType(it.ithChildren(0), "VAR") }

// Remember variable "found"
.sideEffect{ _found = it.ithChildren(0).varToName().next() }
.filter{ isType(it.ithChildren(1), "CALL") }
.filter{ isType(it.ithChildren(1).ithChildren(0), "NAME") }
.filter{ isType(it.ithChildren(1)
   .ithChildren(1), "ARG_LIST") }
...
\end{Verbatim}
   \end{minipage}
   \caption{Generated traversal for the code snippet in
     Figure~\ref{fig:sampleTutorial}.}
   \label{fig:traversal}
\end{figure}

The generated traversals implement a computationally efficient code
analogue detection. For classic code clone detection tools, the
comparison of every subtree of a given AST of $N$ nodes with every
subtree of the same AST costs $O(N^3)$ and can become $O(N^4)$ when
also comparing sequences of trees. Since empirically, a large software
system of $M$ lines of code has $N=10M$ AST nodes, this AST-based
approach suffers a hard scaling problem. There is a solution to this
runtime cost, which involves comparing hashes of AST subtrees instead
of the AST subtrees themselves. Then, the runtime becomes
$O(N)$~\citep[see][]{baxter1998clone}. However, hashing subtrees will
not work with our approach.

In our approach, we are comparing a given source AST $S$ of $N$
nodes, obtained from a tutorial, with a target AST $T$ of M
nodes, which is the code base that should be scanned for code analogues
of $S$. Therefore, $N$ is much smaller than $M$.
For this approach, the runtime is much smaller. More formally, our algorithm starts out with
comparing the first node in $N$ with every node in $M$, until it finds
one that is equal.
This costs $O(M)$.
Then, it will compare the other nodes as well, until all nodes in
$S$ were matched successfully in $T$. Therefore, when a code
analogue is found, we did $N$ comparisons to match it. For subtrees that
are not analogues of $S$, we do 1 to $N-1$ comparisons.
This cost is $O(N)$.
In the worst case, this entails a full check that costs $O(N)$ for
every node in $T$. Thus, the complexity of our approach is $O(NM)$,or
for cases where $N \ll M$ holds, $O(M)$.

Our algorithm can benefit from knowledge about the depth of nodes. Let
$d(x)$ denote the depth of a node $x$, then, comparisons beyond an AST
depth of $d(T) - d(S) + 1$ can be safely avoided, given that nodes at
the depth cannot contain the AST $S$.  However, we typically make use
of small vulnerable code snippets ($N \ll M$), for which the saving in
computational cost is marginal.

Translation of tutorial templates into graph traversals provides us a
means to mine large amounts of open\hyp{}source code for recurring
vulnerabilities.
Since we assume that the attacker is not an insider (who can
stealthily plug our analyses into the back\hyp{}end of a hosting
platform), analysis impinges on the attacker being able to download a
project's source code.
To this end, we have implemented a tool called \emph{GithubSpider} to
facilitate code crawling at scale. In the next section, we briefly
discuss its design and implementation.

\subsection{Spidering Code Repositories}
\label{sec:spidering}

We make a conservative estimate of computing resources that an attacker has at his/her disposal.
We assume that an attacker has access to a modest computing device such as a standard PC, and a DSL broadband connection.
Our threat model lowers the barrier to entry for performing vulnerability discovery at scale.
Additionally, we assume that the attacker makes every effort to be stealthy.
In the context of code spidering, this implies that rate limits imposed by a hosting service cannot be abused.
Thus, we are constrained to spider at a modest speed.
The constraints that we impose on the attacker make our tooling and analysis operate in a real\hyp{}world setting.
Understanding whether an attacker, in spite of these constraints, can use our techniques to find vulnerabilities in open\hyp{}source code is part of our research question.

Although multiple code hosting platforms provide APIs that could be used for spidering, we focus on GitHub in our case study.
GitHub is the most popular open-source collaboration platform, hosting approximately $10.3$ million developers, and over $25$ million projects~\citep{github:frontpage}. In addition to code, GitHub maintains meta-data such as a project's language, creation time, popularity (\emph{stars} and \emph{forks}), and total code size.

Our spidering tool, \emph{GithubSpider}, uses GitHub's REST API to obtain project meta\hyp{}data for a large number of projects.
GithubSpider is designed to be general purpose: It obtains project meta\hyp{}data and applies user\hyp{}defined filters to them, providing the flexibility required to spider code repository classes of interest.
%The design of GitHubSpider is generic enough to crawl other kinds of GitHub repositories.
Thus, GithubSpider can be leveraged towards analysis of a different family of codebases, say \emph{C/C++} code.
We leverage GithubSpider to crawl projects written in the PHP language.
Additionally, we filter projects based on their popularity.
We gauge a project's popularity by the number of stars\footnote{A GitHub user can express interest in a repository by starring it.} it has received.
After narrowing down projects of interest, we leverage GitHubspider to clone (download) projects and their revision history (see Section~\ref{sec:spider}).
% We make our spidering tool available as open-source to encourage
% further research in the area\footnote{\url{https://github.com/tommiu/GitHubSpider}}.

% Without direct access to the hosting platform, the attacker is in need
% of a tool to identify PHP projects, and download them for scanning. We
% explore the feasibility of this approach by developing a general
% purpose spidering tool that makes use of GitHub's REST API. . Access
% to this API is subject to rate limitations that researchers can
% temporarily disable by contacting the GitHub staff, however, an
% attacker with malicious intent is unlikely to be granted this access,
% and hence, we performed our experiments with the default rate
% limitations in place.
% %
% Our tool then proceeds in three steps. First, GitHub is crawled to
% enumerate repositories. Second, for each repository, its meta\hyp{}data is
%  extracted, allowing its key properties to be analyzed without
% requiring downloading of the complete repository. In particular, this
% allows us to select only repositories with a certain popularity
% (indicated by the amount of \emph{stars}) and with a minimum size. We
% make extensive use of these filtering abilities in our evaluation (see
% Section~\ref{sec:spider}). Finally, selected repositories are cloned,
% that is, the code along with its revision history is downloaded.

\subsection{Mining for Vulnerabilities}
\label{sec:mining}

Finally, we automatically mine the downloaded code repositories using the traversals generated from
vulnerable code snippets (see Section~\ref{sec:querygen}). We achieve
this by importing the code into the code\hyp{}mining platform
\emph{Joern}~\citep{Yamaguchi14}. Joern first parses source files to
generate ASTs, and subsequently
imports the ASTs into a Neo4j graph database, which allows us to
efficiently execute graph traversals and collect statistical data about
our analysis.

Scanning the large number of projects obtained from our spider in an
acceptable time frame is challenging task, as graph matching needs to
be performed on all ASTs of all retrieved code
bases. Fortunately, projects can be processed independently, and
therefore, we can distribute scanning across several processes running
in parallel by splitting spidered repositories into groups. This task can also be carried out in a distributed setting.

\begin{algorithm}[t]
\caption{Algorithm to find code analogues using graph traversals.}
\label{ccdetection:pseudocode}

\begin{algorithmic}[1]

\Procedure{MineCode}{$repositories$,
$queries$}
\For{each $repository$ in $respositories$}
        \State create ASTs for $repository$
        % \State derive nodes and relationships files from AST
        \State import ASTs into graph database and start server
        % \State start graph database server
        \For{each $query$ in $queries$}
                \State run $query$ against database server
                \State save statistical data
                \State save code clone data
        \EndFor
        \State stop graph database server
\EndFor
\EndProcedure
\end{algorithmic}
\end{algorithm}

For a given set of queries and a group of repositories,
Algorithm~\ref{ccdetection:pseudocode} summarizes the scanning process that we carry out.
For each repository, the algorithm creates ASTs for
each of its source files and imports them into the graph database. The
graph database server is subsequently started, each query is executed
against the databases, and statistical data such as query execution
time is collected. Upon running all queries on a project, the graph
database server is stopped.

We have implemented the presented algorithm in a tool, that we call \emph{CADetector}, short for code analogue detector.
CADetector takes automatically generated queries (Gremlin traversals), and a PHP project for analysis as input, and returns matching code as output.
Code analogues are presented to a human analyst for review.
In the next section, we present our evaluation methodology and results.

\section{Evaluation}
\label{sec:eval}

% \begin{itemize}

% \item The system employs an AMD Phenom II X4 955 with a clock
%   frequency of 3.2 GHz and a HDD with a writing speed of 78.2 MB/s.

% \item Spidered GitHub projects over a period of two months, collected
%   462,609 PHP repositories. Cloned 84,850. Analyzed 64,415
%   repositories in total. 18 queries, discovered 820 code clones: 117
%   instances that contained at least one vulnerability.

% \end{itemize}

% \begin{table*}[t]
% \centering
% \begin{tabular}{ r | r | r | r | r | r}
% \toprule
% processes & runtime & \# repos crawled & avg \# of repos crawled/hour & avg repository size & total size/hour \\ \hline
% \midrule
% 1 & 275.38 m & 8,372 & 1,824.08 & 268.74 KB & 490.2 MB \\
% 4 & 254.46 m & 21,044 & 4,956.06 & 286.97 KB & 1,422.24 MB \\
% 8 & 231.43 m & 18,997 & 4,925.12 & 251.51 KB & 1238.7 MB \\
% \bottomrule
% \end{tabular}
% \caption{Comparison of GitHubSpider's crawling speeds on a 64-bit
%   Linux distribution with an AMD Phenom II X4 955 (3.2 GHz) CPU and an
%   HDD write rate of 78.2 MB/s. The number of processes run varies and
%   results in different crawling speeds. The optimal number of
%   processes depends on the host's system.}
% \label{eval:spidering_speed}
% \end{table*}

Our evaluation follows a two\hyp{}step process.
First, we select a handful of vulnerable code snippets, obtained from top\hyp{}ranked PHP tutorials, to seed our study.
Second, if we find a match for a vulnerable snippet in our data set as a result of our analysis, we flag it for manual review.
Based on this two\hyp{}step process, we have performed extensive evaluation of our analysis framework.
Our analysis data set consists of 64,415 PHP codebases that have been downloaded using GitHubSpider.
To gauge its feasibility, we have run our analysis against the top 10 PHP codebases on GitHub, in addition to the codebases in our data set.
The following paragraphs describe how seeds were obtained and queries generated (Section \ref{sec:findingtutorials}), the nature of the analysis data set (\ref{sec:spider}), and our analysis results (\ref{sec:code_mining}).

\subsection{Identification of Vulnerable Tutorials}
\label{sec:findingtutorials}

To identify widely read tutorials, we
query the Google search engine using the following set of terms.

\begin{verbatim}
"mysql tutorial"
"php database user"
"php mysql user query"
"php search form"
"php ajax search tutorial"
"javascript echo user input"
\end{verbatim}

For each of these search terms, we manually review the first five results
returned by the search engine. We evaluate each of the tutorials for
SQLi and XSS vulnerabilities by following
established secure programming guidelines by the Open Web Application
Security Project (OWASP), namely, the guide on \emph{Reviewing Code
for SQL Injection}~\citep{owasp:sql_review}, and the \emph{Cross Site
Scripting Prevention Cheat Sheat}~\citep{vuln:owasp_xss_prevention}.
Among the top five results (30 in total), we find 9 tutorials that
contain vulnerable code: 6 tutorials with SQLi, and 3
tutorials with XSS. A snippet from a representative tutorial containing both SQLi and XSS
vulnerabilities is shown in Figure~\ref{fig:sampleTutorial}.

% \begin{figure*}
%   \begin{minipage}[t]{\textwidth}
%      \VerbatimInput[frame=single, numbers=right, fontsize=\scriptsize,
%      commandchars=@`|]{./listings/vulnTutorial.php}
%    \end{minipage}
%    \caption{Identified vulnerable tutorial, containing an SQL injection
%      vulnerability on line $6$, and stored cross site scripting
%      vulnerabilities on line $11$, and $12$.}
%    \label{fig:sampleTutorial}
%  \end{figure*}

\subsection{Query Generation}
\label{sec:query_gen}

% \begin{figure}[t]
%   \centering
%   \lstinputlisting[caption=Tutorial code snippet,label=lst:sqlinj]{listings/SqlInj.php}
%   \centering
%   \lstinputlisting[caption=Vulnerable slice of the snippet,label=lst:sqlinj_relaxed]{listings/SqlInj_relaxed.php}
%    \caption{(Listing~\ref{lst:sqlinj}) Vulnerable tutorial containing a SQL injection vulnerability between lines 4--6; (Listing~\ref{lst:sqlinj_relaxed}) Slice of the tutorial containing the vulnerability.}
%    \label{lst:sampleQueries}
%  \end{figure}

\begin{figure}[t]
  \begin{minipage}[t]{\columnwidth}
\begin{Verbatim}
<?php
// Check for existing user with the new id

@hili`$sql = "SELECT COUNT(*) FROM user WHERE|
@hili`		userid = '$_POST[newid]'";|
@hili`$result = mysql_query($sql);|
if (!$result) {
  error('A database error occurred in 
  processing your submission.\nIf this
  error persists, please contact
  you<at>example.com.');
}
if (mysql_result($result,0,0)>0) {
  error('A user already exists with your chosen
  userid.\n. Please try another.');
}
?>
\end{Verbatim}
  \end{minipage}
  \caption{\label{fig:vulntut}Vulnerable code snippet from tutorial, containing a SQL
    injection vulnerability between lines 4--6 (slice).}
\end{figure}

% \begin{figure}[t]
%   \begin{minipage}[t]{\columnwidth}
%     \VerbatimInput[frame=single, numbers = left, fontsize=\scriptsize]{./listings/SqlInj_relaxed.php}
%   \end{minipage}
%   \caption{\label{fig:vulnslice}Slice of the tutorial containing the vulnerability.}
% \end{figure}

 % \begin{itemize}
 % \item Recurring vulns can occur in two ways: copypaste and similar programming idiom
 % \item We tailor our query gen strategy to cater to both possibilities
 % \end{itemize}

We expect copy\hyp{}pasted code from vulnerable tutorials to result in recurring vulnerabilities in application code.
A more subtle manner in which recurring vulnerabilities may manifest themselves is when developers follow the same (vulnerable) programming idiom(s) presented by a tutorial.
To cater to both possibilities, we generate queries using an exact replica of code present in a tutorial (normal query), and a slice of the tutorial code containing the vulnerability (strict query).
A strict query abstracts only the vulnerable slice of code, whereas a normal query abstracts the entire tutorial.
We use strict queries to identify known vulnerable patterns in web applications, and normal queries to identify code analogues of tutorial code.
Figure~\ref{fig:vulntut} illustrates the difference between code snippets used for generating normal, and strict queries.
Entire tutorial code is shown in the listing with the vulnerable slice highlighted in red.
Lines 4--6 of the tutorial contain a classic SQLi vulnerability: Unsanitized user\hyp{}input from a {\tt POST} variable is used in a MySQL query.
The vulnerable slice contains only these lines, which represents the minimal working snippet containing the vulnerability.

\subsection{Analysis Data-set}
\label{sec:spider}

We leverage GitHub to obtain a large data set of web applications for analysis.
GitHub hosts over 25 million projects written in several programming languages.
Thus, we filter content (PHP projects) that is relevant to us.
Our crawler inspects project metadata (accessible via REST API) to perform the filtering.
GihubSpider, our code crawler implementation, has filtered through a total of 462,069 PHP repositories on GitHub.
Of these, we have downloaded a total of 64,415 PHP codebases for analysis.
These codebases comprise our analysis data set.

We divide our data set into three groups by popularity.
We quantify the popularity of a codebase by the number of times it has been \emph{starred} by users on GitHub.
%In each category, we limit the total size of a codebase.
Our classification results in the following data set partitions:
\begin{enumerate}
\item \textbf{Barely known projects (Not popular).} 42,064 projects that are
  starred at most three times and have a total file size of less than
  3 MB.
%We refer to this data set as \textbf{D1}.
\item \textbf{Projects known by several people (Popular).} 16,037 projects
  that are starred at least four times, but at most nine times, and
  have a total file size of less than 3 MB.
\item \textbf{Popular projects (Very popular).} 6,314 projects that are starred
  at least ten times and have a total file size of less than 3 MB.
\end{enumerate}

\noindent{}GitHub imposes a rate limit of 5000 API requests per authenticated user per hour.
The imposed rate limit proved to be the main bottleneck in downloading repositories.
Although using multiple authenticated accounts for crawling is a simple workaround for removing the bottleneck, we stayed clear of it.
% Table~\ref{eval:spidering_speed} shows the runtime and throughput of GitHubSpider as a function of the number of processes spawned for crawling.

% \subsection{Query generation}
% \label{sec:query_gen}

% Our method for finding recurring vulnerabilities in a codebase hinges on formulating a graph traversal query that can be issued against the analysis target's code property graph.
% We generated queries using \emph{php2gremlin}.
% The queries are based on code snippets found in popular tutorials that have been manually selected using \emph{Google} search.
% %First, we selected a handful of top\hyp{}ranked PHP tutorials using \emph{Google} search.
% Our search sought to elicit tutorial content for common web programming tasks such as interfacing a database with a web application, implementing a search form for a web page etc.
% We selected a total of nine tutorials, all of which were one among the top\hyp{}five hits for a search query.
% We refer the reader to Appendix~\ref{app:search_queries} for a complete list of string literals used for Google searches.

%Given a tutorial code snippet, we generated a query to match the snippet using \emph{php2gremlin} (see \S{\ref{sec:auto_query_gen}}).

% In total, we have automatically generated eighteen queries using nine tutorials (each tutorial accounts for a single strict query and a single relaxed query).

\subsection{Analysis Results}
\label{sec:code_mining}

%\todo{Is presenting query runtime necessary?}

\begin{table}[t]
  \centering
  \begin{tabular}{ p{2cm} p{1cm} p{1.5cm} p{2.2cm} }
  \toprule
  \textbf{Data set} & \textbf{Size} & \textbf{Code analogues} &  \textbf{Vulnerabilities} \\ %\multicolumn{3}{c}{Vulnerabilities}
   & & & (percentage)  \\
  \midrule
%   & & Total & Strict matches & Relaxed matches \\ \hline
  Not popular & 42,064 & 269 & 80 (29.74\%) \\  %& 8 & 261
  Popular & 16,037 & 528 & 35 (6.63\%) \\ %& 0 & 528
  Very popular & 6,314 & 23 & 2 (8.7\%) \\ %& 0 & 23
  \midrule
  Total & 64,415 & 820 & 117 (14.27\%) \\
  \bottomrule
  \end{tabular}
  \caption{Analysis summary for codebases in our data set. In total, 820 code analogues were found which included 117 recurring vulnerabilities. The table shows a break down of our findings in each data set partition.}
  \label{tab:clone_results}
\end{table}

We used auto\hyp{}generated graph traversals (queries) to mine for analogues in our analysis data set.
Discovered analogues were manually reviewed, and vulnerable analogues identified.
Table~\ref{tab:clone_results} shows an overview of the analogues, and vulnerabilities discovered by our analyses.
With under two dozen graph traversals generated from a handful of vulnerable tutorials, we obtained 820 code analogues of which 117 were found to be vulnerable.
We found that string normalization in the AST resulted in non\hyp{}exact matches.
For instance, a query generated from the code snippet {\tt \$var = {\$\_GET['var']}} matched not only its replica but also a seemingly benign snippet such as {\tt \$var = {\$value['id']}}.
Thus, matches returned by the traversals had to be manually validated.
In spite of string normalization, we found that automatically generated queries result in interesting corner cases.
Among non popular codebases, roughly 1 out of 3 code analogues is vulnerable, and on average 1 out of 7 analogues is vulnerable across the entire data set.
Analogues are localized to a small portion of application code, which facilitates manual review of all candidates returned by our analyzer.
%Thus, the moderate number of code analogues returned by our analysis framework do not detract from its usability.

%\paragraph{Analysis speed}
% At the time of writing, we concluded analysis of top\hyp{}10 PHP codebases on GitHub, measured by their starred count.
% Fortunately, we didn't find any code analogues originating in our tutorial data set.
%Figure~\ref{fig:top10_speed} plots CADetector's analysis rate, in terms of lines of code analyzed per second, for all codebases.
%The average analysis rate of CADetector is a little over 100 loc/s, making it fast enough even for codebases with hundreds of thousands of lines of code.

% \begin{figure}[t]
% \centering
% \begin{tikzpicture}
% \begin{axis}[
%   ybar,
%   enlargelimits=0.15,
%   title=,
%   symbolic x coords={laravel, symfony, CI, DP-sPHP, Faker, yii2, composer, WordPress, sage, cphalcon},
%   xticklabel style={rotate=90},
%   xtick={laravel, symfony, CI, DP-sPHP, Faker, yii2, composer, WordPress, sage, cphalcon},
%   ylabel=Analysis speed (LoC/s),
%   nodes near coords,
%   every node near coord/.append style={font=\scriptsize},
% ]
%   \addplot[draw=black, fill=gray] table {plots/top10perf.dat};
% \end{axis}
% \end{tikzpicture}
%  \caption{CADetector analysis speed is shown in terms of lines of code analyzed per second for the top 10 PHP codebases on GitHub.com. The codebases CodeIgniter, and DesignPattern-sPHP have been abbreviated as CI, and DP-sPHP respectively. }
%  \label{fig:top10_speed}
% \end{figure}

\paragraph{Newly discovered vulnerabilities}
We manually verified a total of 117 vulnerabilities in our data set.
Of these, 8 vulnerabilities were replicas of code from a popular SQL tutorial that we found on the first Google results page.
Although all of the 8 vulnerabilities were found among non popular code repositories, the finding shows that ad\hyp{}hoc code reuse is a reality.
We are in the process of notifying the tutorial authors about our findings.
Our hope is that the presented vulnerabilities are fixed in a timely manner, so that developers borrowing code from these tutorials in the future will not inherit the same vulnerabilities in their code.

80\% of the discovered vulnerabilities were SQLi vulnerabilities, and the rest were XSS, and path\hyp{}traversal vulnerabilities.
As shown in Table~\ref{tab:clone_results}, the proportion of vulnerable codebases is higher among low popularity codebases, compared to medium and high popularity codebases.
During manual review, we found a plausible explanation for this disparity in vulnerability density: PHP applications in the moderate and high popularity categories make consistent use of newer, and more secure, MySQL APIs in PHP which are not vulnerable to classic (first\hyp{}order) SQLi attacks.
We found that use of the PDO\_MySQL interface~\citep{pdo_mysql}, and the MySQLi extension~\citep{mysqli} was widespread among these codebases.

\subsection{Discussion}
\label{sec:discussion}

% \begin{figure}[t]
%   \centering
%   \lstinputlisting[caption=SQL injection vulnerability, label=codeclone1]{listings/codeclone1.php}
%   \centering
%   \lstinputlisting[caption=Path traversal vulnerability,label=codeclone2]{listings/codeclone2.php}
%   \centering
%   \lstinputlisting[caption=A potential SQL injection vulnerability, label=codeclone3]{listings/codeclone3.php}
%    \caption{The figure shows three analogues returned by CADetector that match the traversal generated for the vulnerable slice in Figure~\ref{fig:vulnslice}. Although we expected to find only SQLi vulnerabilities, we uncovered a path\hyp{}traversal vulnerability (Listing~\ref{codeclone2}) that matched our query template.}
%    \label{lst:sampleHits}
%  \end{figure}

\begin{figure}[t]
  \centering
  \begin{minipage}[t]{\columnwidth}
\begin{Verbatim}
// SQL injection vulnerability
$questionQuery = "SELECT * FROM Questions
		 WHERE AssnID='$_GET[AssnID]'";
$questionResult = mysql_query($questionQuery);
\end{Verbatim}
   \end{minipage}

   \begin{minipage}[t]{\columnwidth}
\begin{Verbatim}
// Path-traversal vulnerability
$filename = "signatures/{$_GET['id']}.png";
$handle = fopen($filename, "r");
\end{Verbatim}
   \end{minipage}

   \begin{minipage}[t]{\columnwidth}
\begin{Verbatim}
// Potential SQL injection vulnerability
$oldBusSeats = "SELECT businessseats FROM flight
		 WHERE flightNo = $flight[0]";
$busseats = mysql_query($oldBusSeats);
\end{Verbatim}
   \end{minipage}

   \caption{\label{lst:sampleHits}The figure shows three analogues
     returned by CADetector that match the traversal generated for the
     vulnerable slice in Figure~\ref{fig:vulntut}. Although we expected
     to find only SQLi vulnerabilities, we also uncovered a
     path\hyp{}traversal vulnerability that matched our query template.}

\end{figure}

% \begin{itemize}
% \item Queries are actually very effective at eliciting vulnerable code
% \item AST serves as a good level of abstraction
% \end{itemize}

Manual review of the code analogues and vulnerabilities returned by our analysis framework suggests that, graph traversals are very good at eliciting vulnerable snippets in a large amount of code.
Firstly, our approach ensures that the code analogues that we find are small snippets of code, typically spanning under 10 lines of code in our dataset.
Figure~\ref{lst:sampleHits} shows three code analogues returned by CADetector for a query originating from the vulnerable code snippet in Figure~\ref{fig:vulntut}.
All three analogues span two lines of code and can be quickly assessed by a human analyst.

Secondly, we find that the abstraction that we choose (AST augmented with data\hyp{}flow information) is robust.
For example, we discovered a path\hyp{}traversal vulnerability in the process of mining a codebase using a SQLi query.
Indeed, both vulnerabilities share the same syntactic structure of code: A tainted PHP variable is used in a security\hyp{}sensitive PHP function call (\code{mysql\_query} and \code{fopen}).
However, since our abstraction does not convey information about taintedness of data, our analysis returns analogues where vulnerabilities may need to be manually verified.
The third analogue shown in Figure~\ref{lst:sampleHits} serves as a demonstrative example.
The analogue contains a potential SQLi vulnerability.
It is vulnerable if the PHP variable {\tt \$flight} reads from a tainted variable (say, an attacker\hyp{}controlled {\tt \$\_POST} variable).
This needs to be manually verified in our setup.

\paragraph{Analysis runtime}
Our timing measurements showed that our analysis is fast for even relatively large codebases.
For the top 10 PHP code repositories on GitHub, CADetector analysis runtime varies between 19 seconds (for the \emph{laravel} project, 777 lines of code) and 53 minutes (for \emph{symfony}, 209 thousand lines of code).
Our evaluations suggest that CADetector is a fast analogue detector for codebases with hundreds of thousands of lines of code and above.

\section{Related Work}
\label{sec:relatedwork}

Our work touches upon two distinct problems: finding similar code, and flagging vulnerabilities in source code.
In the following, we contextualize our work in both domains.
\\

\paragraph{Code clone detection}
Despite modern software design processes and state of the art programming environments, real\hyp{}world software development accommodates ad\hyp{}hoc code re\hyp{}use.
In their seminal work on code clone detection, Baxter et al.~\cite{baxter1998clone}, citing earlier work~\cite{lague1997assessing, baker1995finding}, state that 5-10\% of source code is duplicated in large software projects.
The initial motivation for code clone detection was that ridding software of seemingly redundant code might achieve a performance gain.
Thus, traditional code clone detection tools seek code replicas in a single codebase, or a set of codebases with the same provenance.
This has guided the design of several code clone detection tools~\citep{balazinska2000advanced, baxter1998clone, ducasse1999language, kamiya2002ccfinder, komondoor2001using, krinke2001identifying, Gabel2010, Jiang2007, kamiya2002ccfinder, Li2006, Lanubile2003, bulychev2009evaluation, juergens2009clonedetective, hummel2010index, jia2009kclone, koschke2006clone}.

Recent research~\citep{Jang2012,Pham2010} has shown that code clones pose a more serious threat: Vulnerabilities in cloned code get propagated but their fixes do not.
ReDeBug~\citep{Jang2012} flags unpatched code clones by finding replicas of a known vulnerable snippet in an OS distribution.
Like earlier proposals on code clone detection, ReDeBug flags clones within codebases of similar provenance, because of which it may look for exact matches.
In contrast, we cannot always expect to find borrowed code from an external source \emph{as is}: Developers typically adapt tutorial code for their own end.
This subtle difference precludes the use of hashing functions to measure similarity of code in our work.
Instead, our queries attempt to recognize the structure of vulnerable code.

%Vulnerable snippets are extracted using \emph{diff} files that may be obtained from a software version control system.
%There are three noteworthy differences between ReDeBug and our proposal.
%First, we seed our vulnerability search using
% but also for flagging recurring vulnerabilities that manifest as code clones~\citep{Jang2012, Pham2010}.
% Our work is closer to the latter goal: Find recurring vulnerabilities that manifest as code clones.
% However, a noteworthy difference between frameworks such as ReDeBug~\citep{Jang2012} and ours is that, we can't always expect recurring vulnerabilities to manifest as exact replicas of tutorial code (copy\hyp{}paste vulnerabilities).
% Developers tend to adapt code from tutorials for their own end, making simple changes such as function/variable renaming.
% This precludes the use of hashing functions in our work.

Yamaguchi et al.~\citep{Yamaguchi2012} propose a machine\hyp{}learning
based method for \textit{extrapolating} (i.e., finding other instances
of) known vulnerable code patterns that are manually specified.
Our work is closer to theirs in that we employ structural code fragments
(such as AST fragments) to drive the search for vulnerabilities.
Having said that, a notable difference is that Yamaguchi et al. perform
computations on a code abstraction (specifically, a vector space).
In our work, the query for a similar code snippet is concretized in the
form of graph traversals.
Moreover, we automatically generate vulnerability patterns from code snippets.

\paragraph{Vulnerability discovery}
Since we use static program analysis in discovering vulnerabilities, we shall restrict our discussion to prior work in this domain.
The dynamic nature of web programming languages, such as PHP and JavaScript, has made static analysis of web applications a challenging task.
Researchers have approached vulnerability discovery in PHP code as a static taint analysis problem: Detect the flow of untrusted user input into a security sensitive sink.
Pixy~\citep{Jovanovic2006} is a static analysis tool that flags XSS, and SQLi vulnerabilities in PHP codebases.
In the same vein as Pixy, Xie et al.~\citep{Xie2006} present a summary\hyp{}based static analysis algorithm to discover security vulnerabilities in PHP code.
Our proposal is not \textit{another} vulnerability scanner for PHP code.
Rather, our techniques provide a means to draw inferences about unsafe coding practices among web application developers.
Considering that web applications are ultimately user\hyp{}facing programs that handle sensitive data, our study is timely.

% Since web applications are user\hyp{}facing and are capable of handling private and security\hyp{}sensitive data, our study
% In summary, prior research has focused on
% Both proposals cater to the problem of finding an unknown vulnerability in web applications.
% Research on finding recurring vulnerabilities in web programming languages is sparse.
% To the best of our knowledge, we are the first to propose a static code scanner for recurring vulnerabilities in a PHP codebase.
%In contrast, our work is premised on the intuition that web applications might already contain vulnerable code snippets from publicly accessible tutorials.
%Thus, our analysis framework is intrinsically query\hyp{}driven.
%There are many static analysis tools for checking PHP code.
%PHP Mess Detector (PHPMD), RIPS, and so on.
% \paragraph{Code clone detection}
% PMD's Copy-Paste Detector (CPD) (Cite?) tool supports PHP code clone detection.
% It uses string matching to find clones, not AST fragments.

\section{Limitations and Future Work}
\label{sec:limitations}

% \begin{itemize}
% \item Techniques more general than implementation
% \item Leaves room for expansion to other languages such as C/C++
% \item Our queries don't match up to vulnerabilities whose description is meta\hyp{}syntactic
% \item More systematically leverage popular tutorial content for Vulnerability discovery
% \end{itemize}

A limitation of our study is that our prototype restricts the
evaluation scope of recurring vulnerability detection to PHP application code; that is, we cannot say that programmers employing other languages are similarly
prone to copying from tutorials. Moreover, we restrict our analysis to
open-source code, and thus, the possibility exists that the practice
of copying from tutorials is particularly prevalent in the open-source
world and less common in closed-source environments. Exploring these
questions is left for future work.

For the detection of code analogues, we employ an approach that allows
the names of identifiers to be changed, but is otherwise strict about
the code it matches. For example, if additional statements are
introduced in between statements of a seed, we do not detect the
corresponding code as a clone. This is a deliberate design
choice. Although it may result in the discovery of fewer tutorials,
the identified code is more similar to that contained in the tutorial,
and therefore, more likely to have been copied from it.

Our \emph{approach}---formulating graph traversal queries from code
snippets, and issuing these queries in a code mining
system---is generic enough to be decoupled from the specifics of a
programming language. Thus, our analysis techniques can be
incorporated into existing code analysis platforms such as
Kythe~\citep{kythe}, Joern~\citep{Yamaguchi14}, and
Frapp{\'e}~\citep{frappe}. Systematically leveraging popular tutorial
content from the Web to seed vulnerability discovery is an avenue for
future research. For instance, portals such as Google Trends can be
queried to obtain high\hyp{}value seeds for vulnerability discovery.

\section{Conclusion}
\label{sec:conclusion}

% \begin{itemize}
% \item Briefly restate hypothesis and evidence for and against it
% \item Restate original motivation
% \item Evaluate the novelty of this paper, and provide a reassessment based on its findings
% \end{itemize}
Developers routinely consult programming resources as software is written.
Although formal documentation such as language and API reference manuals provide detailed guidance, tutorials on the Web are as easily available and are more succinct.
The lure of quick actionable advice makes tutorials an appealing reference for developers.
We find that tutorials are not only ubiquitous on the Web but also very popular, consistently appearing in the first Google results page.
Several tutorials betray a lack of understanding of secure coding practices advocated by well\hyp{}regarded online communities such as OWASP.
In our large\hyp{}scale case study, we find over 100 vulnerable code snippets in application code that are syntactically similar, and in 8 instances identical, to tutorial code.
These findings corroborate our hypothesis that vulnerable tutorials can be used to seed large\hyp{}scale vulnerability discovery.
They also suggest that there is a pressing need for code audit of widely consumed tutorials, perhaps with as much rigor as for production code.

We show that the syntactic structure of code can be used to infer similarities between code snippets of different provenance.
Because syntactic analysis is relatively lightweight, it is fast enough to mine a large number of differently\hyp{}sized codebases for recurring vulnerabilities.
Our large\hyp{}scale study is a testament to the efficiency of our proposal.
This also means that there is low barrier to entry for performing vulnerability discovery at scale.
Even an adversary with access to modest computing resources may be seen as direct threat to the security of software in the large open\hyp{}source landscape.
Although our study provides a single data point for an objective assessment of both our adversarial strategy and the connection between tutorials and real\hyp{}world software, there appears to be promise in the applicability of our techniques to other application classes.

% In this paper, we have presented a large scale study of open-source
% PHP projects, which shows that open-source projects suffer from
% vulnerabilities introduced by copying and pasting snippets of code
% from popular but flawed tutorials.  As an immediate practical
% consequence, we are able to identify 117 tutorial-induced
% vulnerabilities in open-source projects. We achieve this by spidering
% code from the open-source collaboration platform GitHub and processing
% it using automatically generated graph traversals for the code mining
% platform Joern.

% Our findings show that the  identification of vulnerabilities in
% tutorials is a worthwhile approach for attackers interested in
% bootstrapping large-scale vulnerability discovery. From a defender's
% point of view, this suggests that code snippets in tutorials should be
% audited for vulnerabilities with the same rigor as production code, as
% failure to do so can result in a proliferation of vulnerable code
% across the open-source landscape.

% In this paper, we have presented a novel approach for large\hyp{}scale
% vulnerability discovery. Instead of analyzing an application in
% isolation, we identify vulnerable code patterns in documentation that
% programmers usually consult. Then, we try to find a match for the
% identified patterns in application code using code property graphs and
% lightweight static analysis.  Using this approach, we have
% demonstrated that an attacker with limited resources can scan large
% amounts of publicly accessible code for recurring vulnerabilities.

\bibliographystyle{splncs03}
\bibliography{master}

% \appendix
% \section{Google search strings}
% \label{app:search_queries}
% We issued Google search queries to obtain a list of top\hyp{}ranking tutorials.
% Our queries contained the following string literals : {\tt php database user}, {\tt php mysql user query}, {\tt php search form}, {\tt php ajax search tutorial}, {\tt javascript echo user input}, {\tt mysql tutorial}.
\end{document}